\documentstyle[psfig]{elsart} 

\begin{document}
\begin{frontmatter}
\title{The mass distribution in the innermost regions of
Spiral Galaxies.
}

\author 
{ Charu Ratnam $\&$ Paolo Salucci}  
\address{
International School for Advanced Studies, SISSA, Via
Beirut 2-4,
I-34013 Trieste, Italy 
}

\begin{abstract}

We use high-spatial  resolution   ($\sim 100\ pc$) 
rotation curves of 83 spiral galaxies  to investigate  
the mass distribution of  their innermost  kpc. We  show  
that, {\it in  this region}, the luminous matter  completely accounts for  the
gravitational potential and no dark component is  required. The derived 
I-band disk mass-to-light ratios ${\mathcal{ Y}}_I $   agree  well with  those  
obtained  from population
synthesis models and  correlate   with color in a  similar way.  We
find strict  upper  limits of $\sim 10^7 M_\odot$  for  the masses of compact bodies 
 at the center of   spirals,  ruling out that these systems
  host the remnants of the quasar activity.

\end{abstract}

\end{frontmatter}

\section {Introduction} 

There is  increasing direct evidence that, at the centers of
bulge-dominated galaxies there reside Massive Dark Objects
(MDO), which are probably the   remnants of the engines that once
powered the  QSO phenomenon (Ho, 1998; Kormendy \& Richstone, 1995). In fact, 
virtually every hot 
galaxy hosts a MDO/BH with a mass ranging from $\sim
10^8 M_\odot$ to  $2\times 10^{10} M_\odot$,  similar to those
related  to the  QSO phenomenon.  
For disk galaxies, the situation is different and much more
uncertain. 
A direct determination of the central mass has been obtained
only in 
very few cases which include our own Galaxy where a black hole with a mass 
$2\times 10^6 M_{\odot}$ has been discovered (Ghez \etal 1999,  Genzel 1998; 
see also Salucci et al 1999 for other few cases). 
Remarkably, these masses do not exceed $10^7  M_\odot$;
however, the lack of detections of very massive objects
($M_{MDO} >10^8  M_\odot$) cannot be ascribed to
observational biases.
In fact,  the   rotation curves (RC),  in great number available 
down to $r_{in }\sim 100 \ pc $,  could easily  
expose central bodies with masses  of the order of $\sim 10^{8}
M_\odot$,  given that  usually  $V(r_{in}) \sim \ 10\  km/s$, 
or equivalently, the stellar mass inside $100 pc $ barely reaches $10^7
M_\odot$. On the other hand,  the  same RC analysis  determining the
MDO mass, obtain also the mass distribution  of  the   innermost regions.
This  is particularly important in the case of low luminosity 
DM-dominated spirals (see Persic and Salucci, 1990)
for which it is  generally difficult 
to disentangle the   disk  component from the whole bulk of    
gravitating  matter. In any cases, let us stress that  to proper  model the 
 region where the luminous matter  dominates is indispensable
 to infer  the structural properties of dark matter.

%%%%%%%%%%%%%%%%%%%%%%%%%%%%%
%%
%
%%%%%%%%%%%%%%%%%%%%%%%%%%%%%
%%
%%%%%%%%%%%
\begin{figure}
\vspace{12cm}
\includegraphics{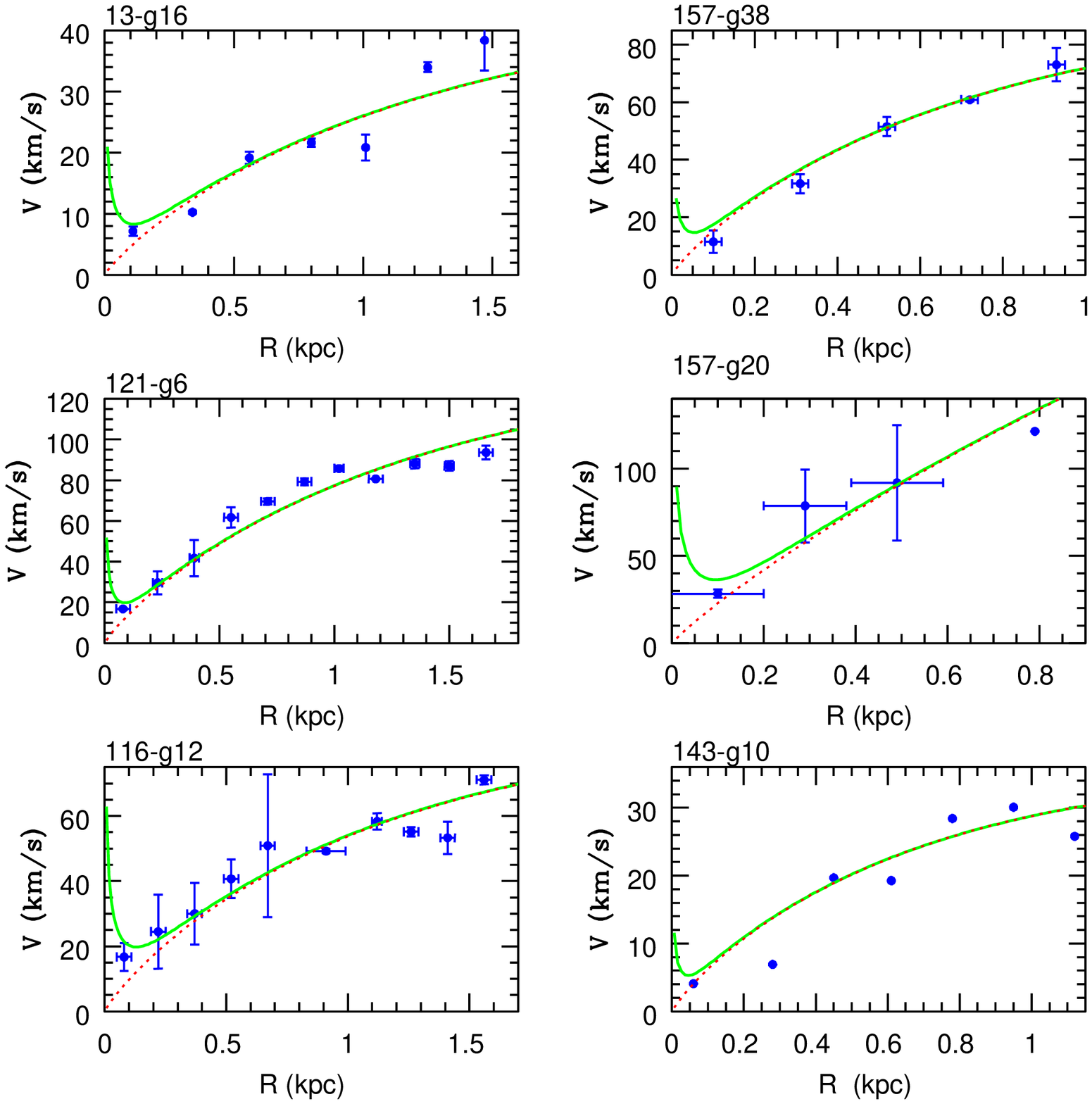}
\vspace{10cm}
\includegraphics{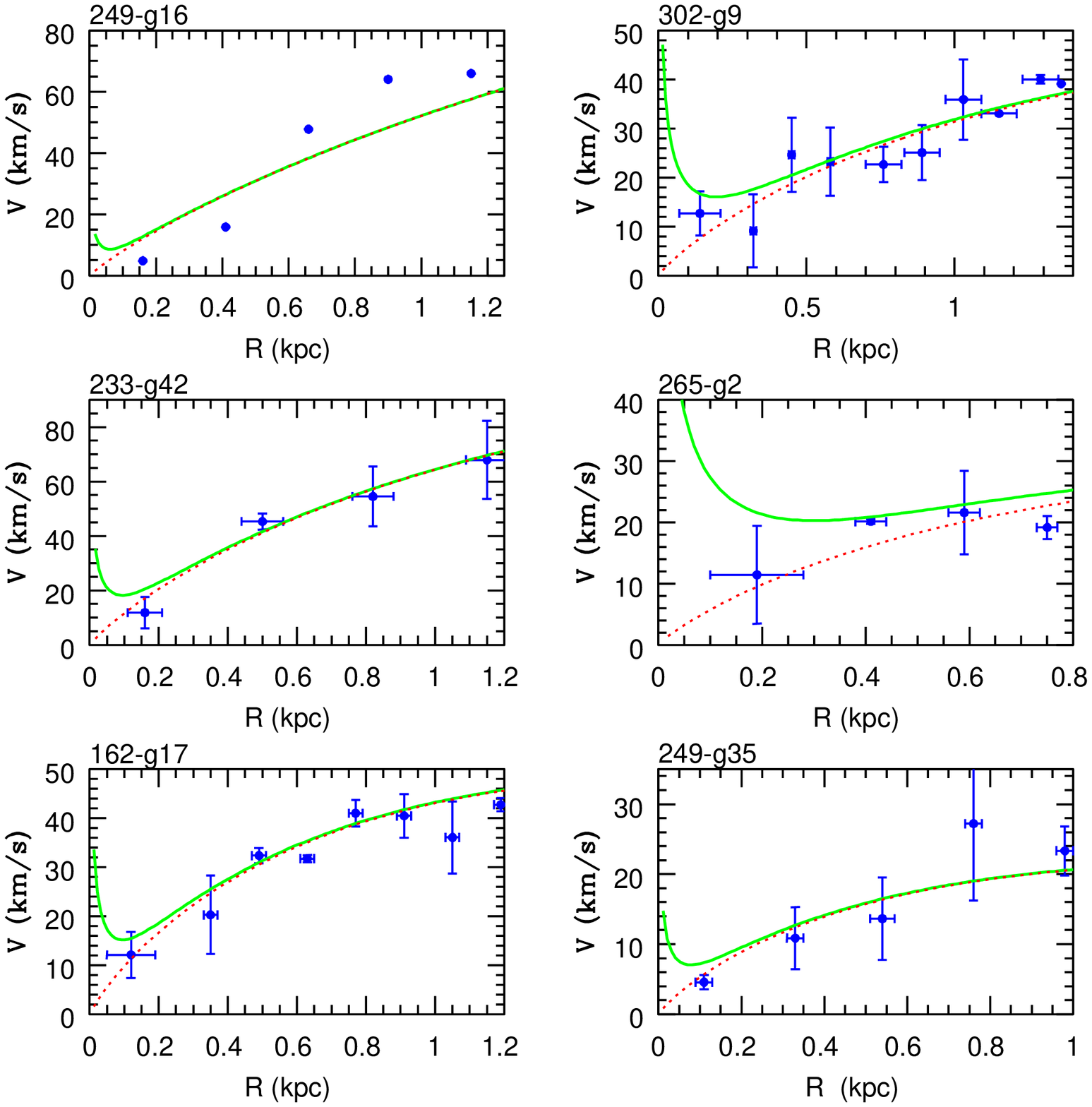}
\end{figure}
%%%%%%%%%%%%%%%%%%%%%%%%%%%%%
%%
%%%
%%%%%%%%%%%%%%%%%%%%%%%%%%%%%
%%
%%%%%%%%%
\begin{figure}
\vspace{12cm}
\includegraphics{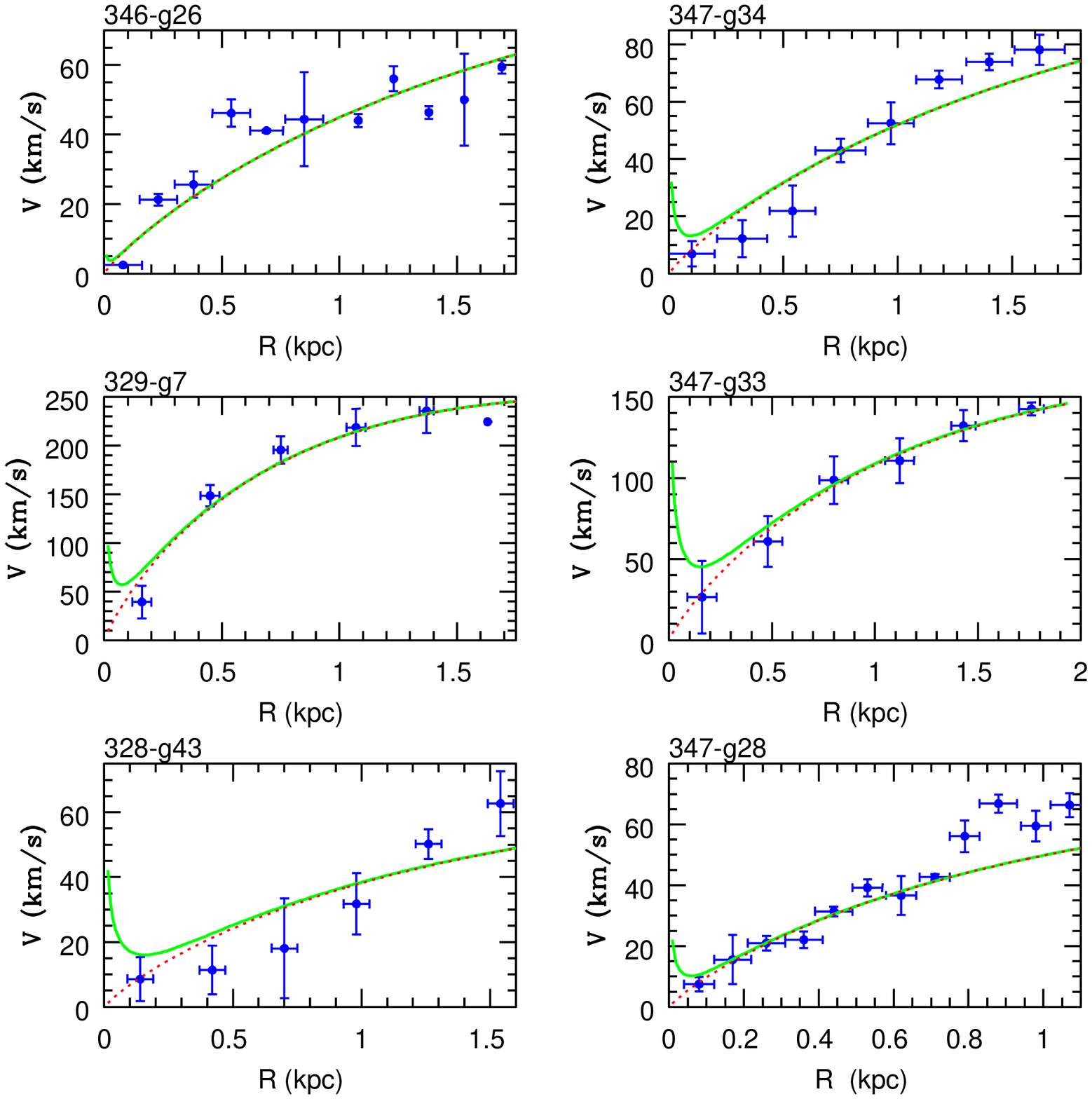}
\vspace{10cm}
\includegraphics{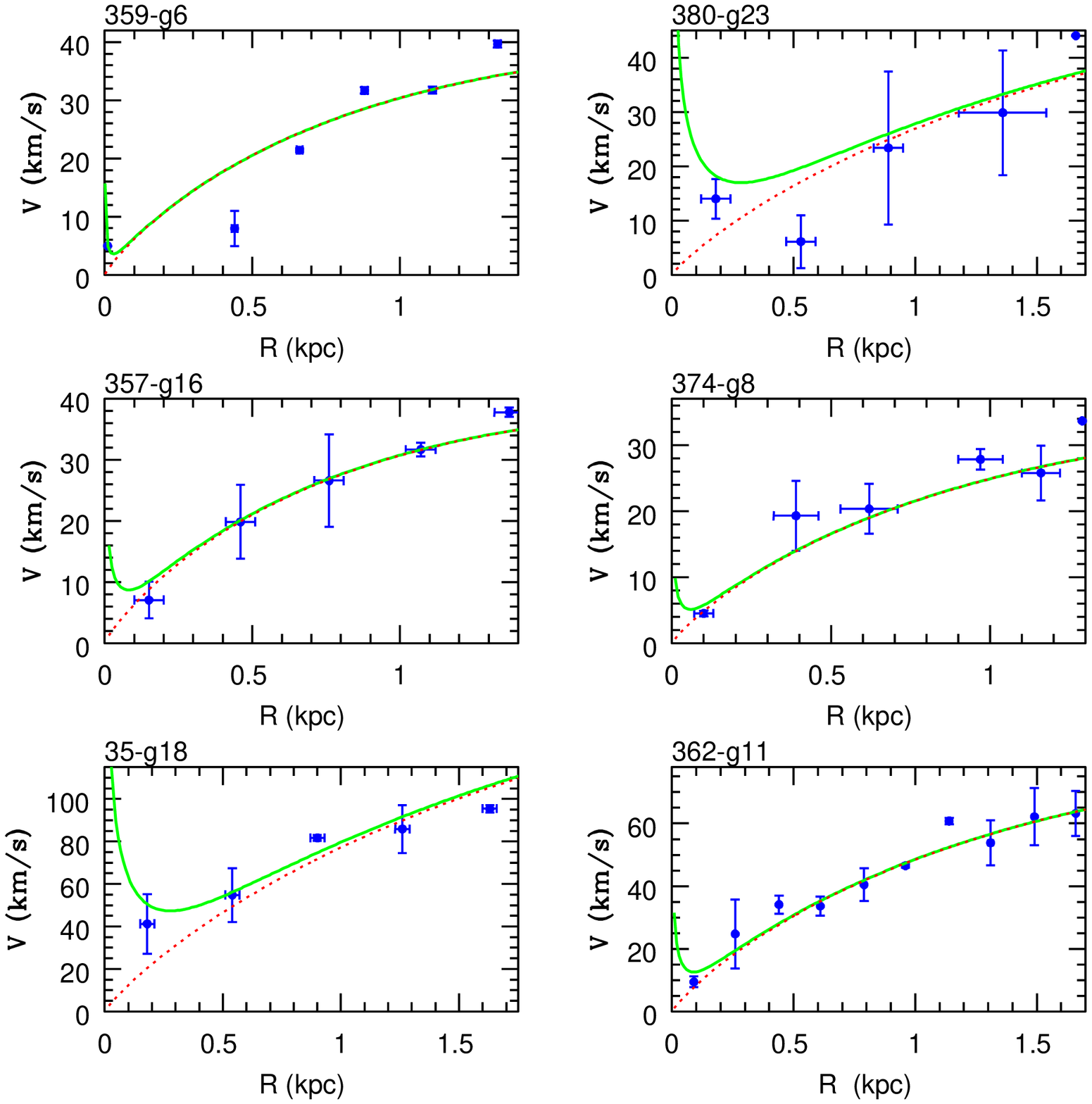}
\end{figure}
%%%%%%%%%%%%%%%%%%%%%%%%%%
%%%%%%%%%%%%%%%%%%%%%%%%%%
%%%%%%%%%%%%%%%%%%%%%%
\begin{figure}
\vspace{12cm}
\includegraphics{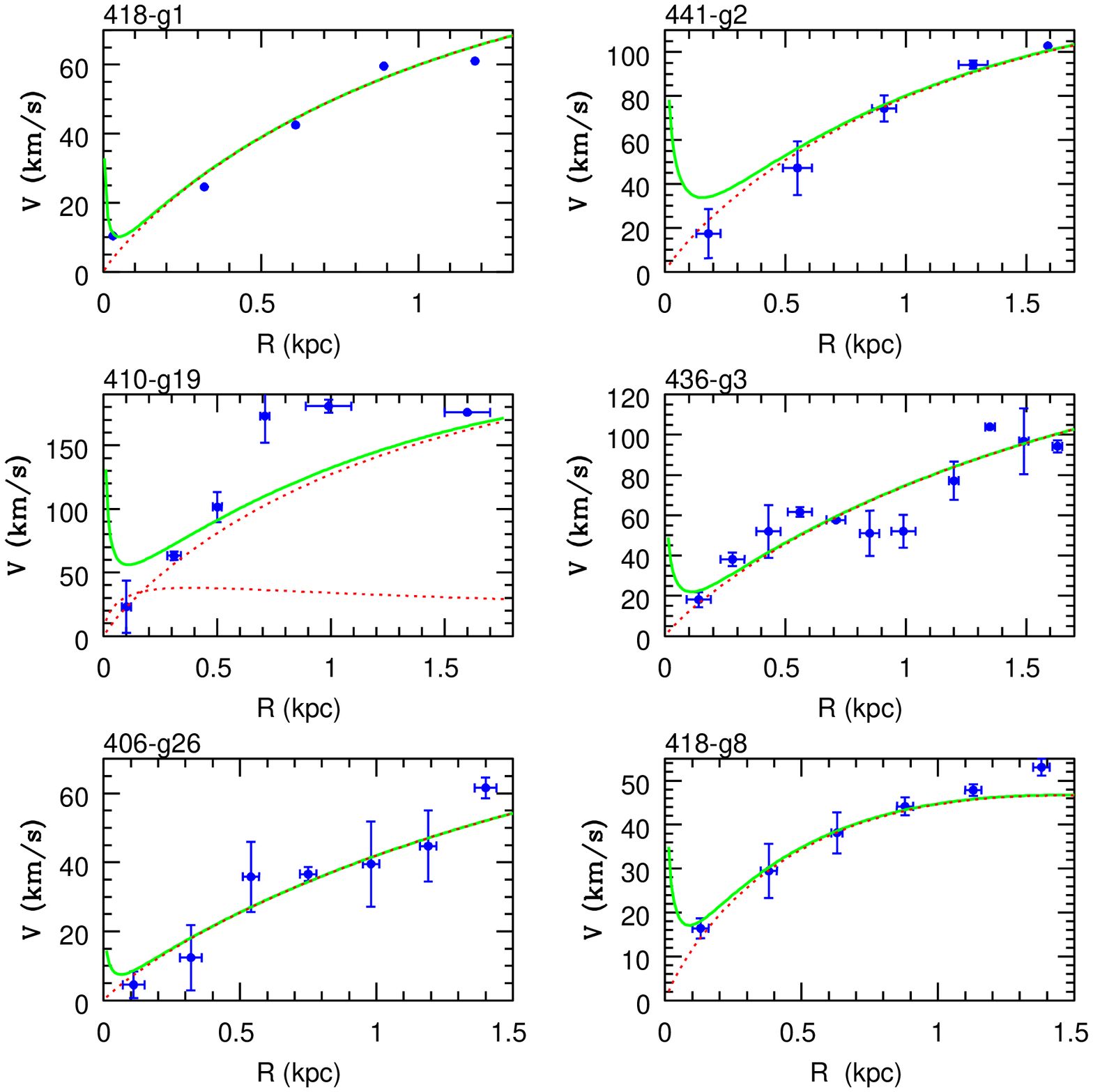}
\vspace{10cm}
\includegraphics{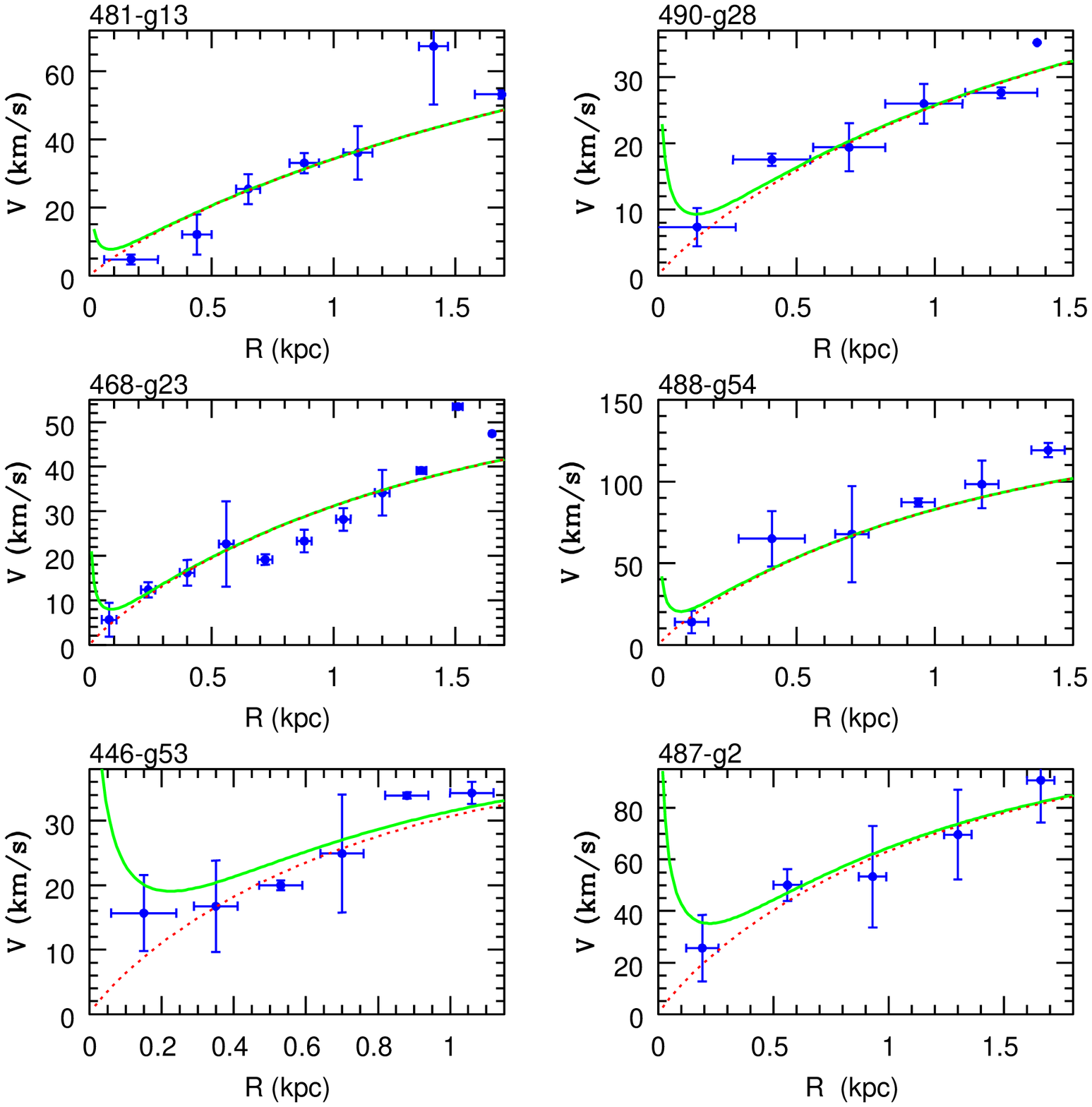}
\end{figure}
%%%%%%%%%%%%%%%%%%%%%%%%%
%%%%%%%%%%%%%%%%%%%%%%%%%%%%%
%%
%%%%%%%%%%%%%%%%%%

\begin{figure}
\vspace{12cm}
\includegraphics{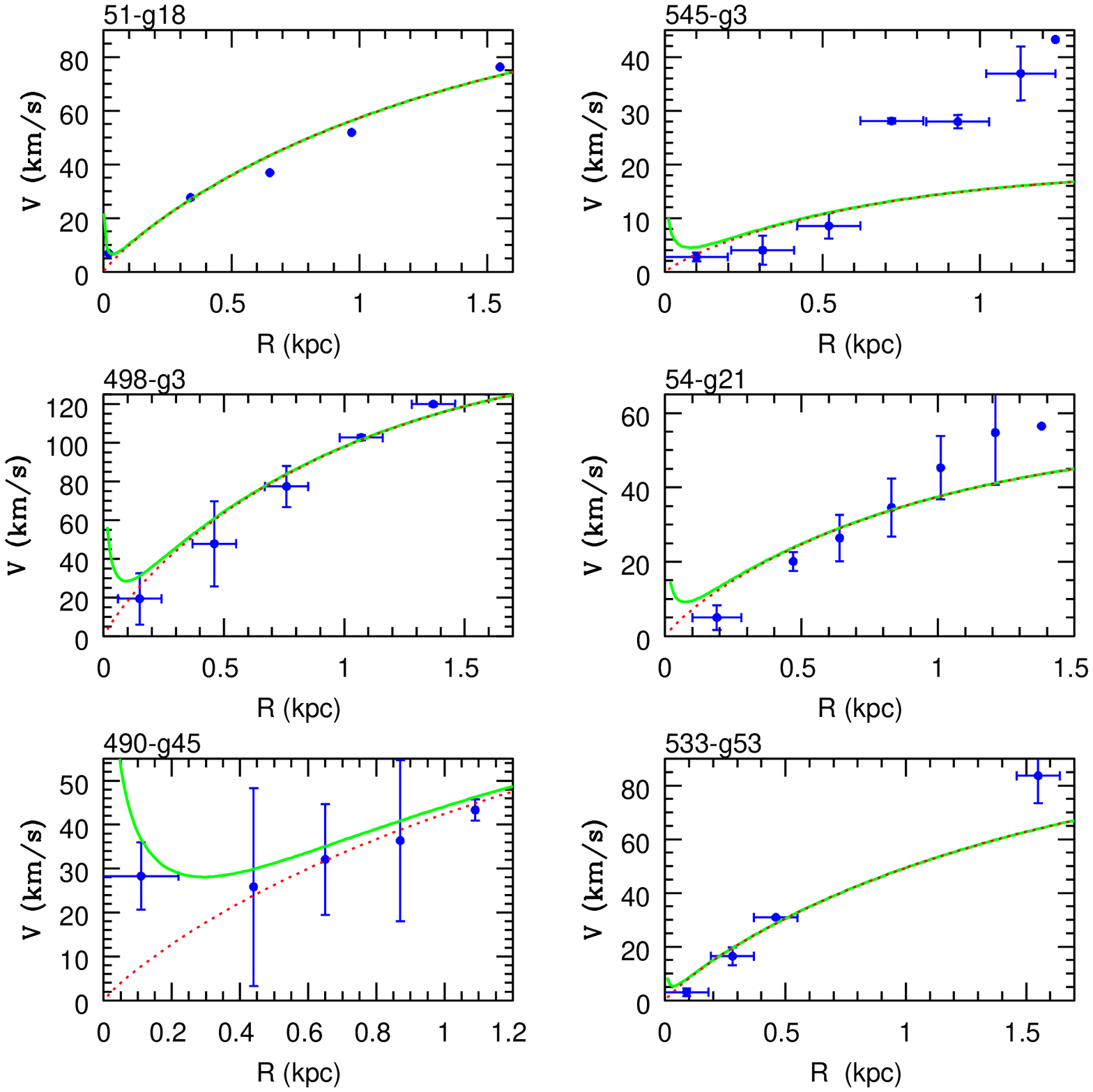}
\vspace{10cm}
\includegraphics{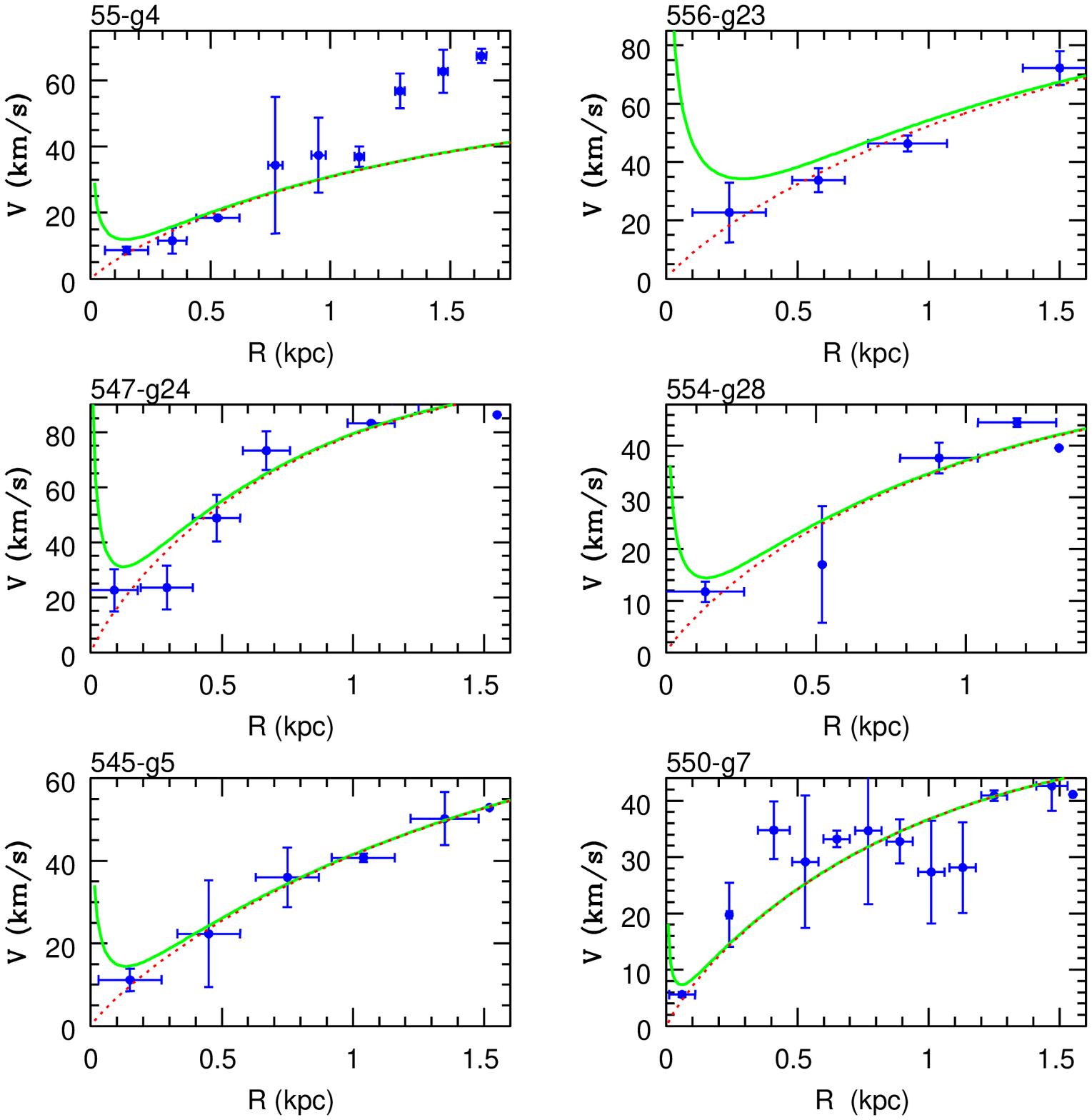}
\end{figure}

%%%%%%%%%%%%%%%%%%%%%%%%%%%
%%%%%%%%%%%%%%%%%%%%%%%%%%%%%
%%
%%%%%%%%%%%%%%%
\begin{figure}
\vspace{12cm}
\includegraphics{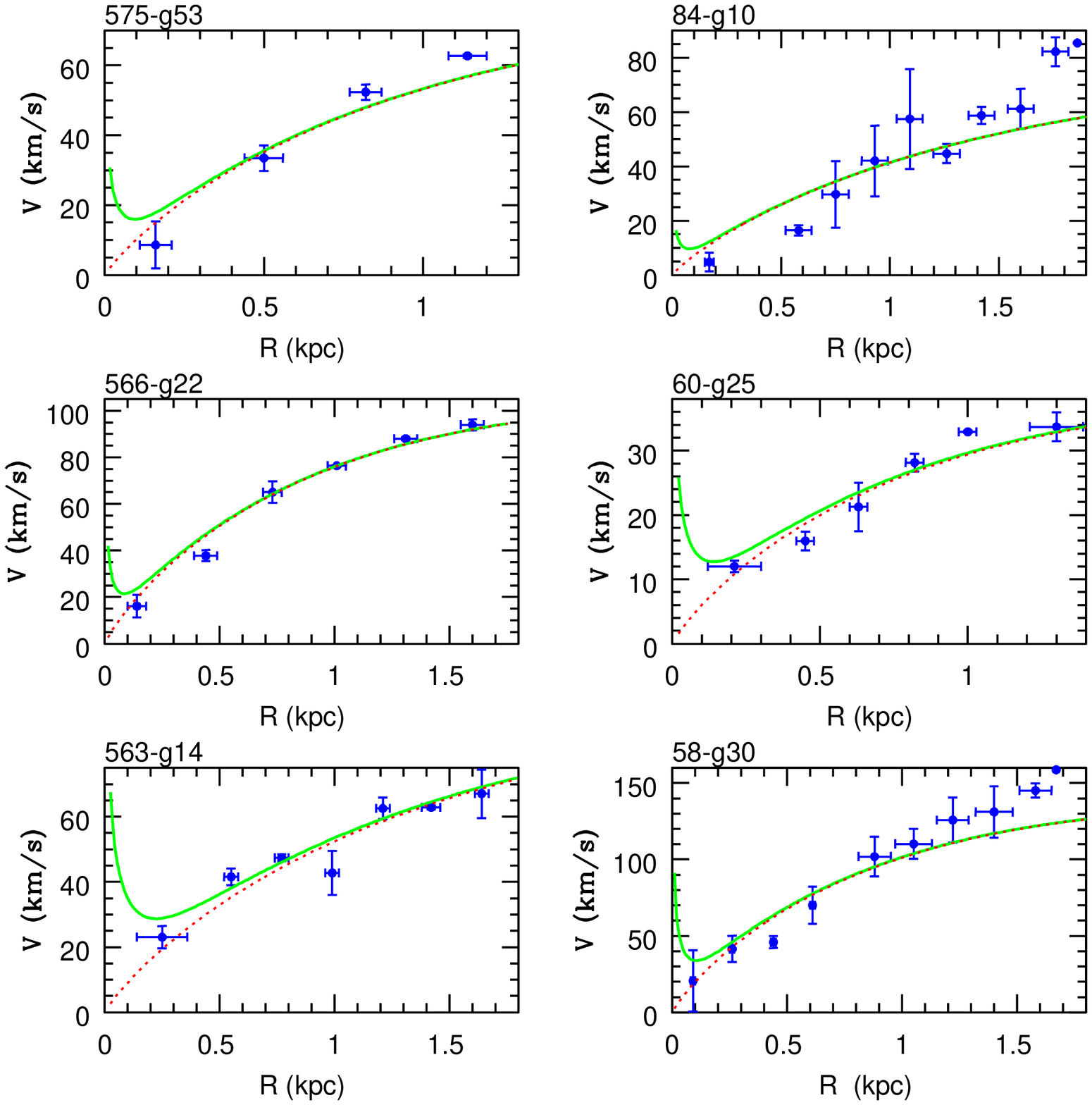}
\vspace{10cm}
\includegraphics{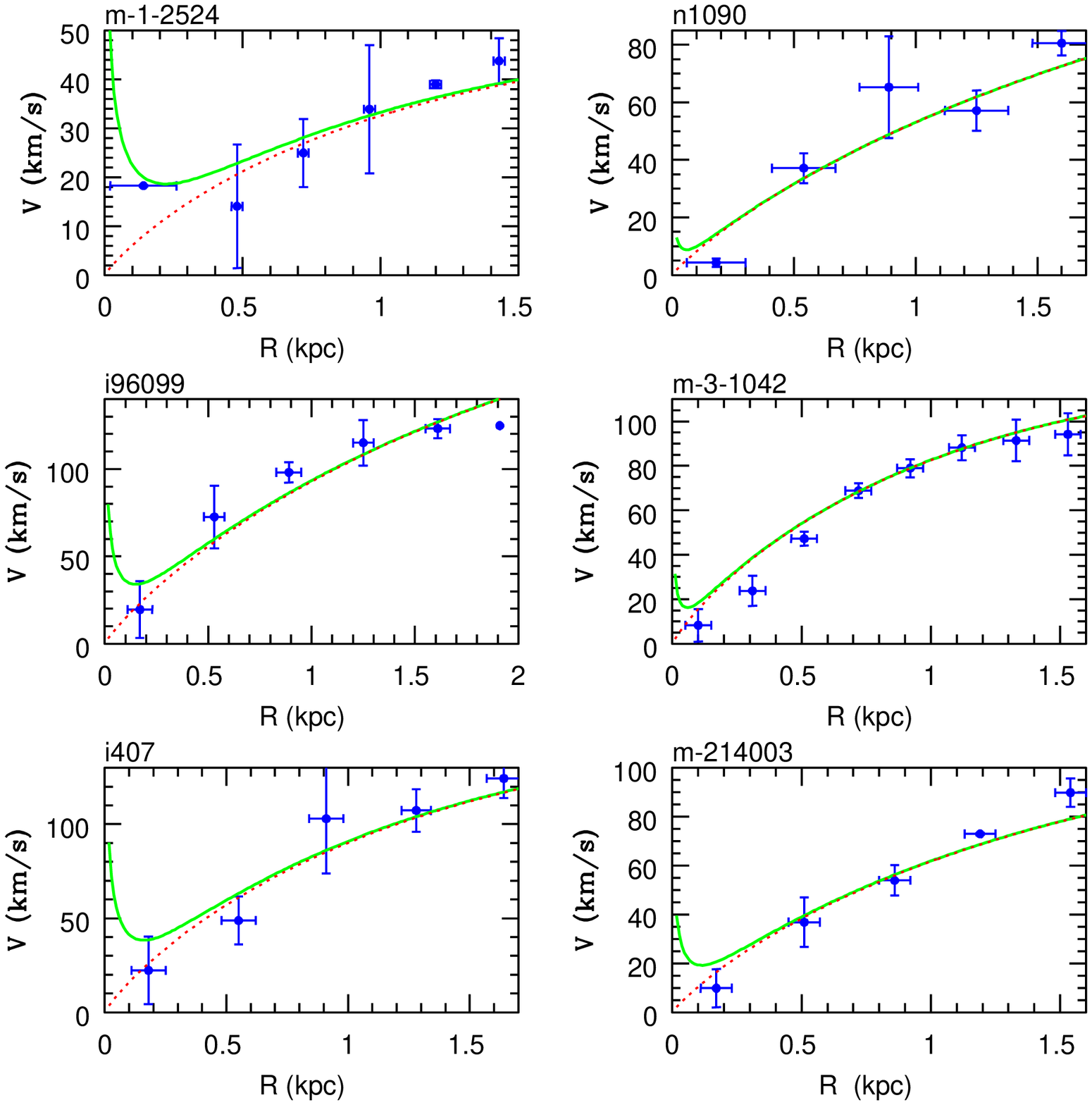}
\end{figure}
%%%%%%%%%%%%%%%%%%%%%%%%%%%%%
%%
%%%%%%%%%%%
%%%%%%%%%%%%%%%%%%%%%%%%%%%% 
\begin{figure}
\vspace{12cm}
\includegraphics{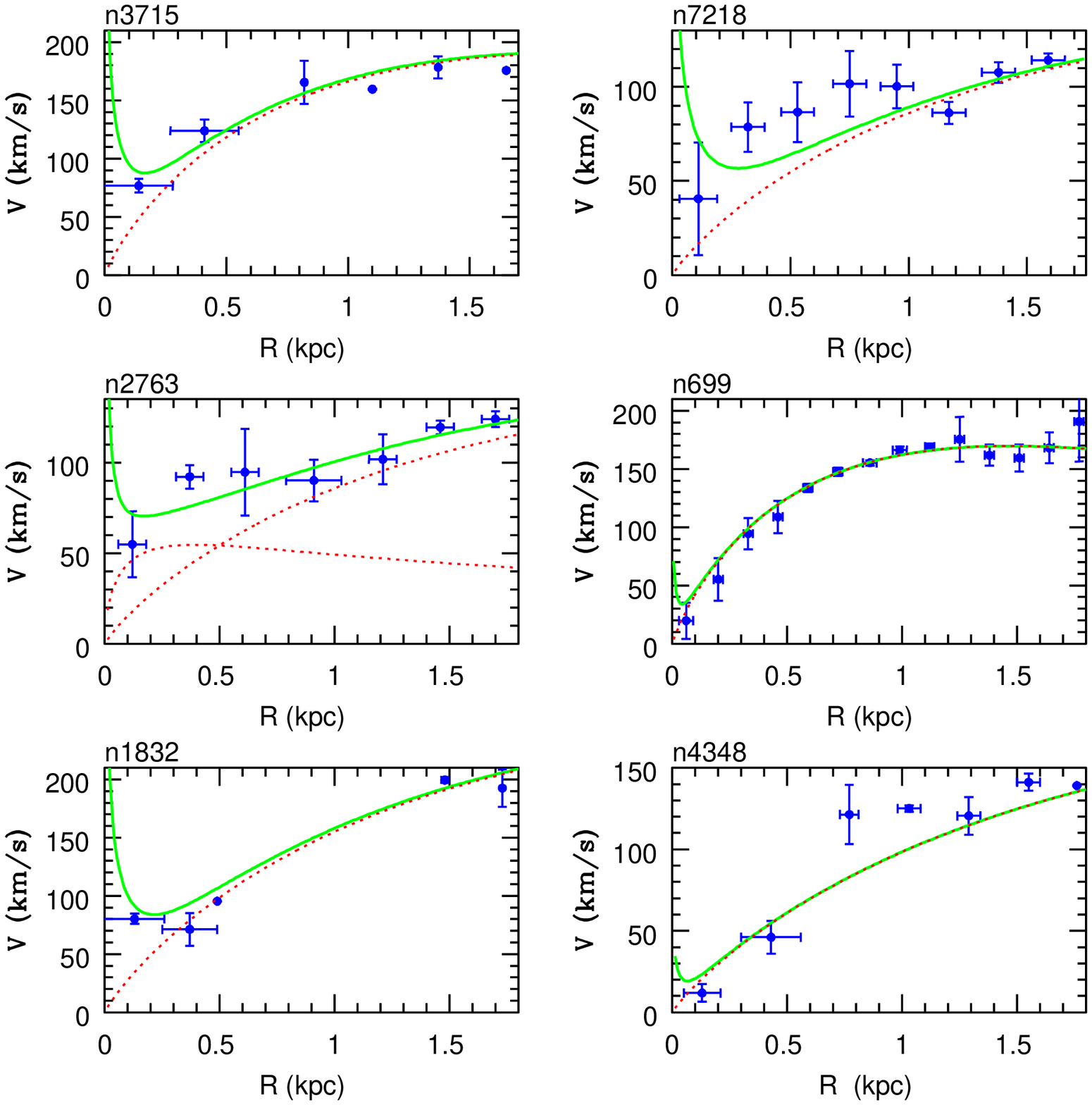}
\vspace{10cm}
\includegraphics{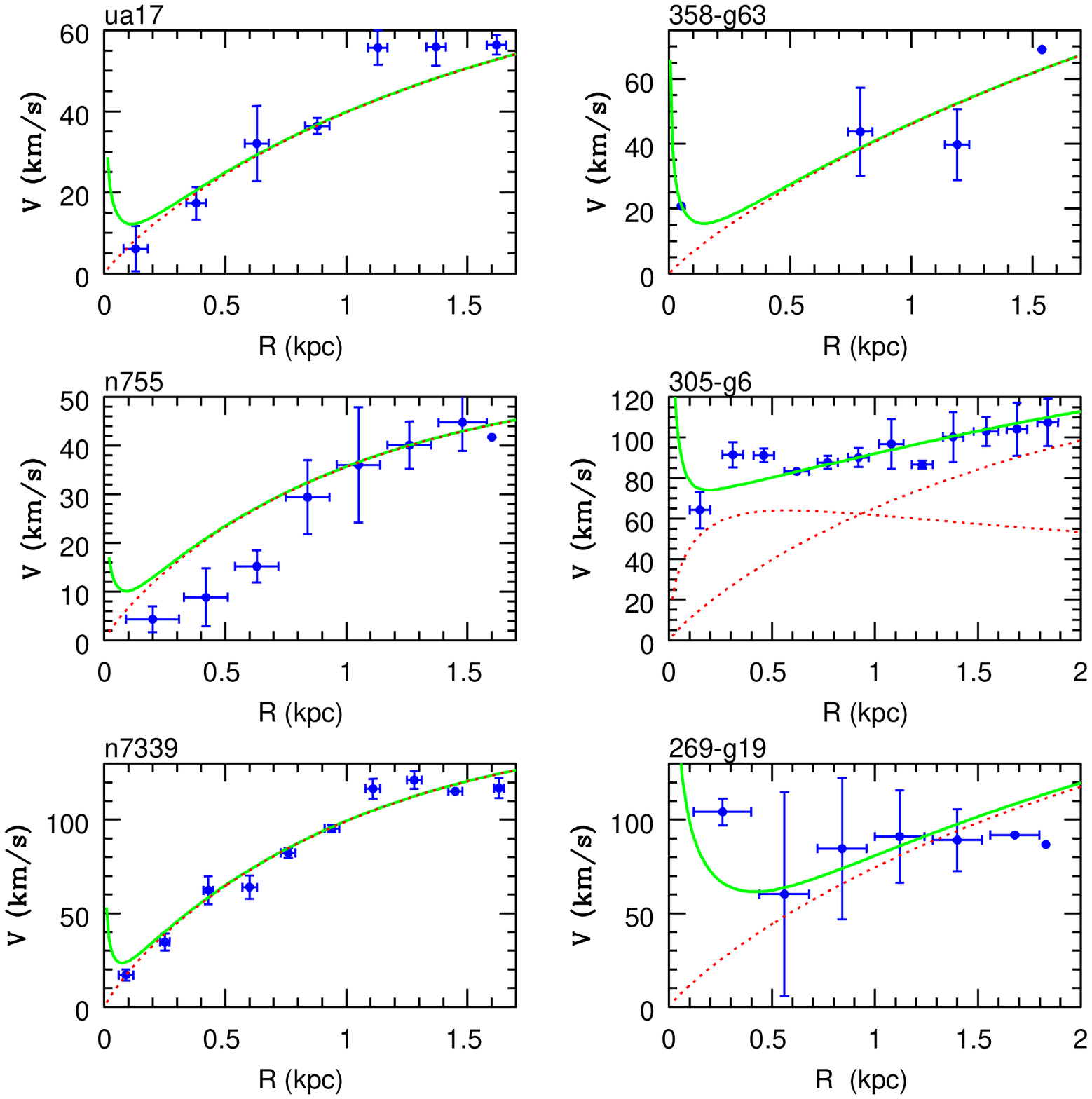}
\end{figure}
%%%%%%%%%%%%%%%%%%%%%%%%%%%%%
%%
%%%%%%%%
%%%%%%%%%%%%%%%%%%%%%%%%%%%%%
%%
%%%
\begin{figure}
\vspace{12cm}
\includegraphics{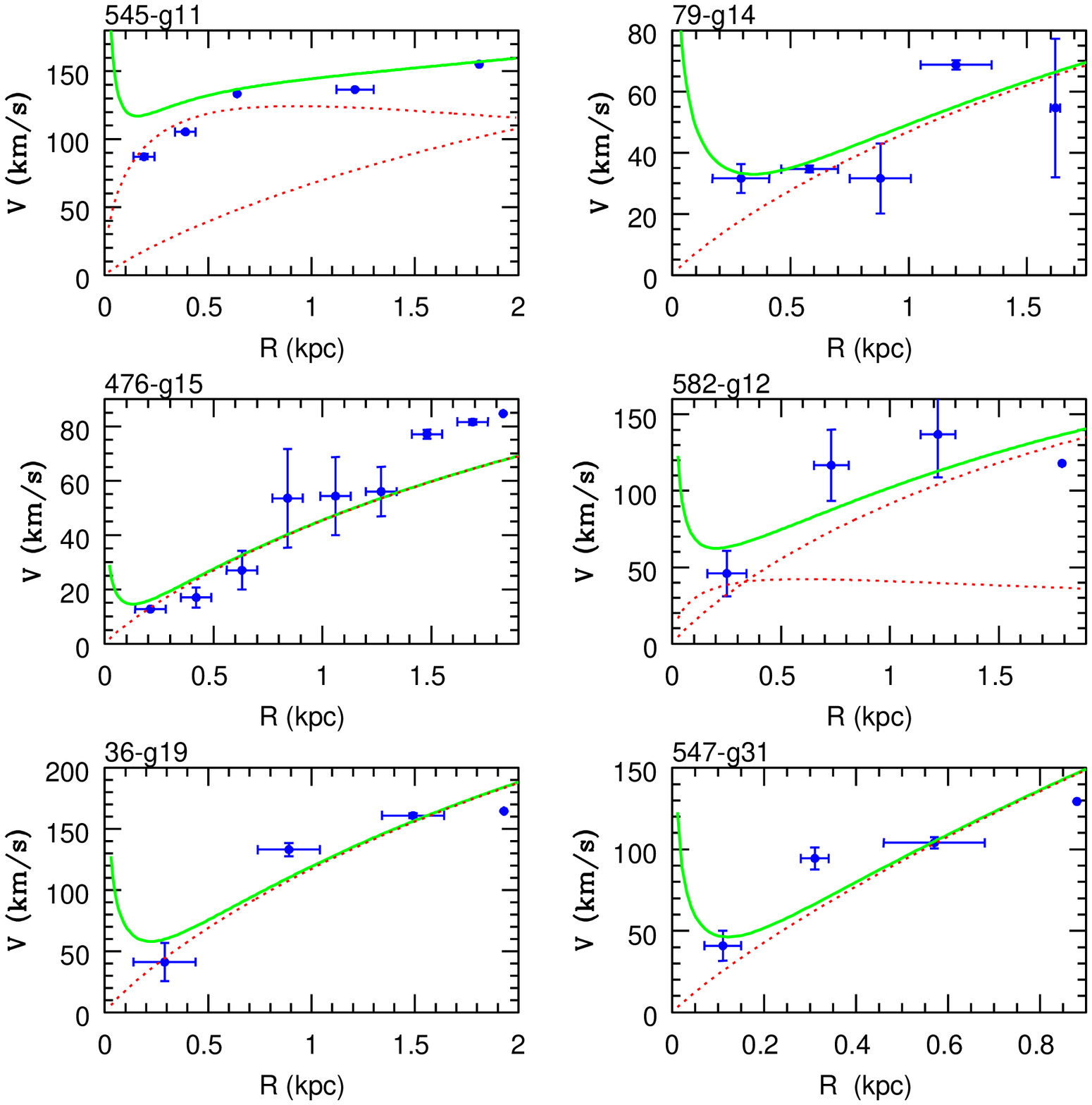}
\vspace{11.4cm}
\includegraphics{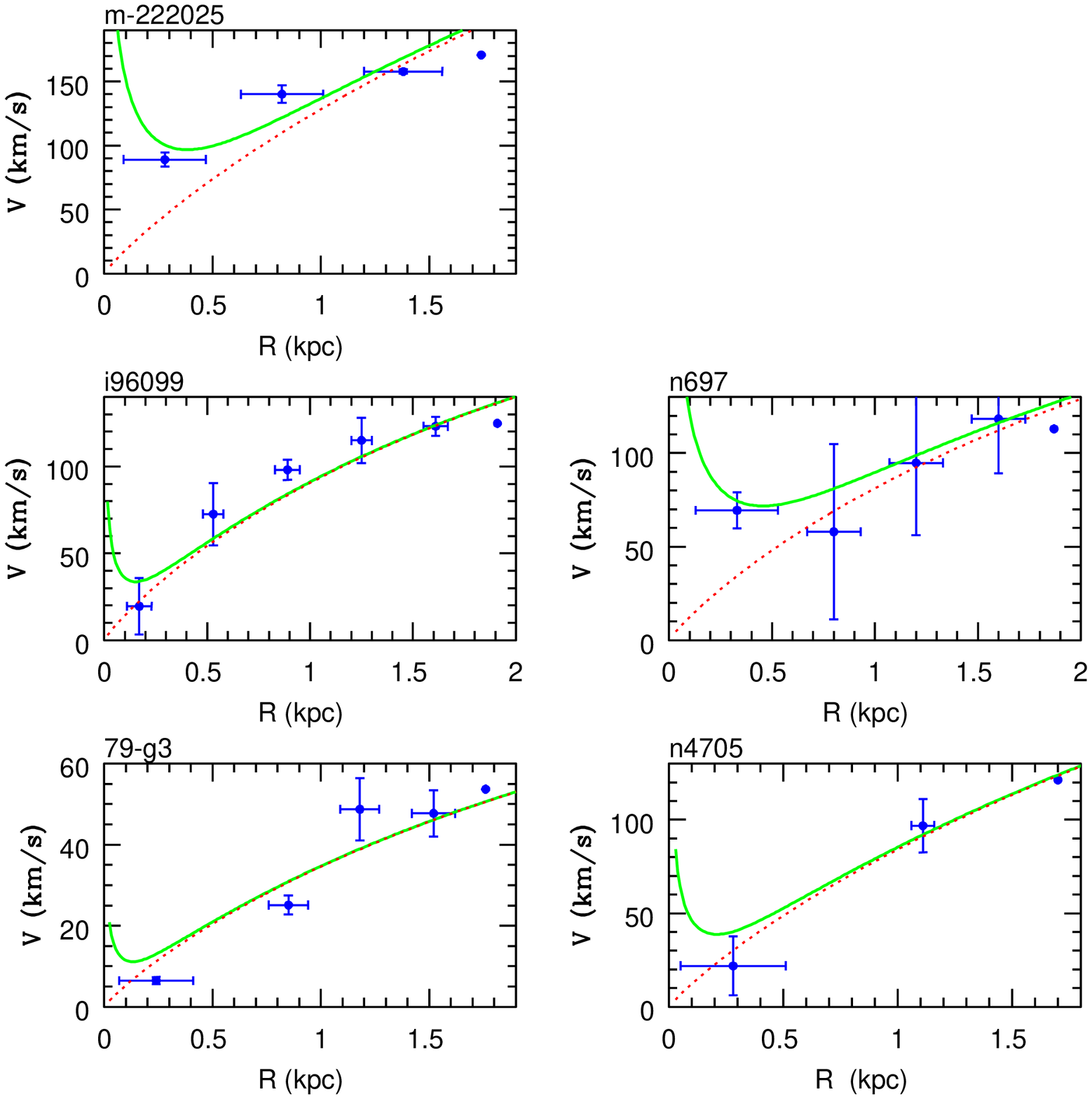}
\caption{Mass models of the  83 objects  of
sample B.
The dashed lines indicate the OD  model (in a few cases
a bulge has been included) while the solid line indicates the disk+BH 
model.}

\end{figure}
%%%%%%%%%%%%%%%%%%%%%%%%%%%%%
%%
%%%%%%%
%%%%%%%%%%%%%%%%%%%%%%%%%%%%%
%%
%%
The aim of this work is to study the  innermost
kinematics of spirals in order to   {\it a)} investigate   the 
distribution of matter in relation to that  
of light
{\it b)} set stringent  upper limits on the  MDO masses. 
 The plan of the paper is the following: in section 2 we  
analyse a sample drawn from 
the 967 rotation curves  (RC's)
of Persic and Salucci (1995), in section 3 we
derive and discuss 
the disk properties and  in sections 4 and 5 we derive 
upper limits on the MDO/BH  masses and 
discuss the results.  
In this paper we assume $H_0=75 \ km/s \ Mpc^{-1}$,  and
take, as the reference 
magnitude in the B-band,  $M_{*}=-20.5$,  which translates
to $M_*-21.9$ 
in the I-band (e.g. Rhee, 1997). All luminosities correspond
to the I-band unless otherwise specified.

\section{Inner Rotation Curves of  Late Types Spirals and
mass modeling}

Recently,  about a thousand
rotation curves of spiral galaxies (PS95), 
tracing the kinematics inside the central kpc, have been available.  
Among these, about one  hundred have  at least one
measurement  $<200\ pc$ and several $<500-1000 \ pc$. 
For these objects the detection of a  MDO of $\sim 10^8 M_{\odot}$ or more,
if present, 
is guaranteed and so is the determination of the  
stellar disk mass, provided that in this region the stellar disk is the 
 major mass component. 
 
Therefore, we select  from PS95 the largest  sample of  
rotation curves  equally distributed in magnitude interval and 
of sufficient high-quality  for our purposes,  by setting the
following 
criteria: {\it i)} each RC has at least  4  measurements inside
a radius of  $340 (1-{(M_I+15)\over {9}})\ pc\sim (350-500)\ pc$, 
{\it ii)} for each RC,  the innermost data  is situated
at a radius $r_{in}<100(1  -{( M_I+15)\over 18})\sim (100-150)\ 
pc$.  Let us notice that  to restrict the  criteria  
reduces the number of RC's without improving their already
high quality that  however rapidly  decreases as they are relaxed.
The  rotation curves resulting from the selection  are 
generally smooth, axisymmetric and with negligible non
circular perturbations.  It is evident that there is a very
good agreement
between the rotation fields  
on the receding and the approaching  arm (see PS95). The 
sample (Sample B in Salucci et al., 2000) has  83  objects, 
well mixed in luminosity and almost equally distributed 
between  Sb-Sc and Sd-Im Hubble types.
  
We aim to  reproduce the RCs by a mass model 
featuring: ({\it i)} a Freeman disk of  length-scale $R_D$,
(derived in  PS95)   
and  mass $M_D$, which    contributes    to the circular velocity as:
$$                       
 V^2_d(R)={1\over 2} GM_D/R_D \ x^2
(I_0K_0-I_1K_1)_{x/2}
\eqno(1)
$$
where $I_n, K_n$ are  modified Bessel functions and
$x=R/R_D$, and 
({\it ii)} a MDO which contributes
to $V(R)$ as:
$$
 V^2_{MDO}(R)=GM_{MDO}/R
 \eqno(2a)
 $$  
where the MDO mass is: 
 $$
 M_{MDO}=f G^{-1} V^2(r_{in}) r_{in}
 \eqno(2b)
 $$ 
with $0<f\leq1$. 

In  Fig.(1) we compare the velocity  
data  with the best-fitting  mass models in  the cases of 1)  
$f=0$ (only disk, OD) mass model and of 2) $f>0$ 
(disk+ black hole) mass model (in practice: $f\simeq 1$).
 As result, we find no central-body-dominated
(CBD) rotation curve, i.e.   no RC shows the Keplerian 
fall-off expected for a CBD RC. On the contrary, 
rotation curves are strikingly close to those predicted  by the OD model, i.e.
by a self-gravitating exponential thin disk with radially constant
mass-to-light ratio. 
The results of the mass modeling are:  in  57 objects 
the OD  velocity curve   accounts for  the rotation data  in a {\it excellent} way and it  
lies within the  data error bars (for $R<R_{IBD}$ see below).
In 14 cases\footnote{349-G6, 38-G23, 35-G18, 157-G20, 436-G3, 468-G23, 547-G24,
550-G7, N4348, 358G-63, 582-G12, 547-G31, I96099, N4705}the luminous  matter   
accounts for  the rotation curve in a {\it satisfactory} way: data and 
model differ utmost by $2\sigma$ and no evidence for an additional component emerges. In 9 cases\footnote
{347-G84, 328-G43, 249-G16, 84-G10, N755, 79-G14, M222025, N7218, 249-G16}  the
OD fits are {\it reasonable} expecially  if  we consider that the corresponding  
RC's   have non-negligible  internal dispersion  and/or some   asymmetry.
 The low-luminosity
galaxy 545-G3  is dominated by dark matter at any radius. 
Finally,  in 2 cases; 346-G26,  410-G19,  the OD  model   reproduces  the data
with some difficulty. 

A simple inspection of the results shows that in  every galaxy  only
for  $R\geq R_{IBD}$\footnote{Inner Baryon Dominance}the dark component  
begins to 
 significantly contribute to the total gravitational 
potential. We then confirm the picture  of  Salucci \& Persic (1999) and 
Salucci etal,
(2000)
according to which, the luminous matter dominates an 
innermost region of spirals  of size   
$R_{IBD}$. This truly baryonic  scale  turns out to be a  
function of galaxy luminosity by  ranging from  $0.5R_{D}\sim 0.5 kpc$
to $2R_{D} \sim 30 kpc$ along the luminosity sequence
Such a transition can be seen, e.g. in the RC of 
 55-G4: OD curve reproduces the data  only  out  to
$R=R_{IBD}\sim 0.5 kpc$.
More in general, the best-fit  models (see Fig 1) clearly show
that,  with the exception of the least  luminous  galaxy, at  
$R\sim 0.5 \ kpc$, the DM
contribution to the RC  
lies  below the detection
threshold. 
On the other hand, in
order to become the  major component at  a $2-3$ disk  scale-lenghts,     the dark
mass, negligible at $0.5 kpc$   must {\it strongly} increase with radius.

We conclude by stating  that the OD  mass models   reproduce the innermost kpc of
the rotation 
curves extremely well:  on this scale,   there is
{\ it no hint of  a    
dark component}.

\section {Disk properties}
%%%%%%%%%%%%%%%%%%%%%%%%%%%%%
%%
%%%%%%%%%%%%%%%%%%%%%%%%%%%%%
%%
%%%%%
\begin{figure}
\vspace{8.9cm}
\includegraphics{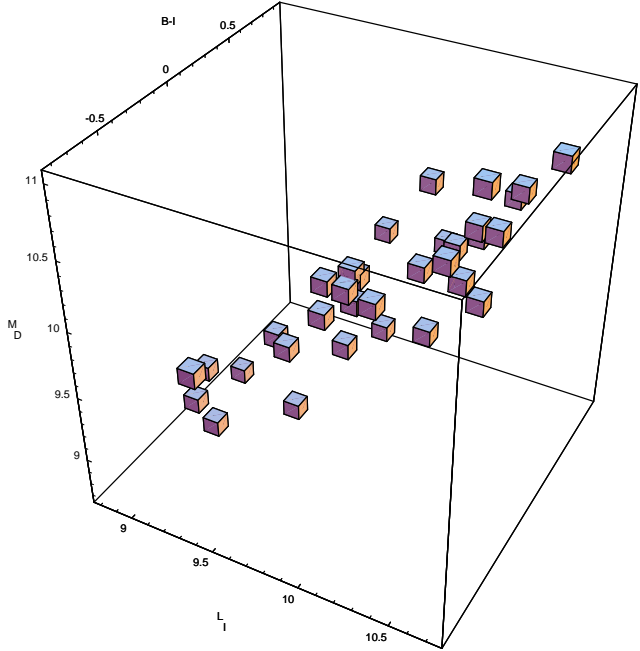}
\caption{The 3D space of  galaxy properties.  
$M_D-(B-I)-L_{I}$.}
\label{fig:fig2}
\end{figure}
%%%%%%%%%%%%%%%%%%%%%%%%%%%%%
%%
%%%%%%%%%%%%%%%%%%%%%%%%%%%%%
%%
%%%%%%%
Disk masses are derived from eq (1); 
 Although the uncertainties on the {\it disk}
mass-to-light ratio values must include   model-fit uncertainties,
photometry
errors, and  uncertainties on the assumed inclination and distance
(the latter present also for
   Tully-Fisher distances).
  The  uncertainty  budget  can be  roughly estimated as:
$\delta M_{D}/L_{I} \simeq  0.4 M_{D}/L_{I}$.

The present sample of DM-free mass models is a  most
suitable one for  investigating the disk masses in relation with
the photometric properties of galaxies. In detail, 
this is done with a sub-sample of Sample B comprising 28 objects for which   
$I$-band luminosities and 
$B-I$ colors are available.  
In Table 1 we 
report  best-fit disk masses,  color, mass-to-light ratios $M_{D}/L_{I}$ 
and luminosity for these objects. 
B-I color are good indicators of stellar mass-to-light ratio (Vazdekis
et al 1996).
In  Fig.2 we the  show  ${\mathcal Y}_I$ vs. $B-I$ colors
alongside with   with the same relation
 predicted by population synthesis models. We find
$$
{\cal Y}_I \simeq 0.63\times  \Big({1.6 L_B\over {L_I}}\Big)^2
\eqno(3a)
$$
in good agreement with Vadzekis et al   
\begin{figure}
\vspace{10.cm}
\includegraphics{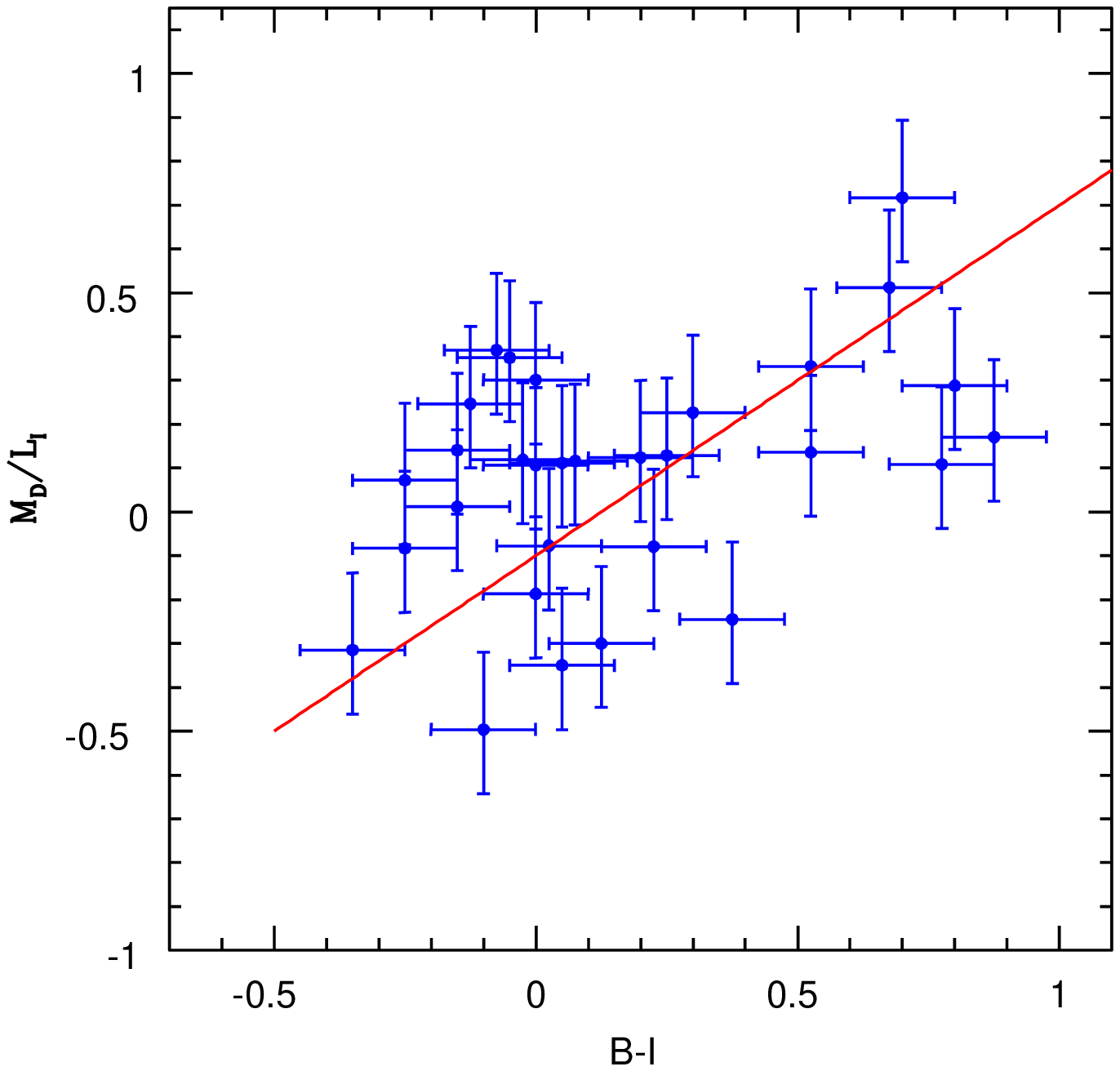}
\vskip -0.8truecm 
\caption{Logarithm of the mass-to-light ratio  in the I band as a function of 
the B-I color.
The solid line is from Vazdekis etal. 1996. The uncertainty in  
B-I is 0.1 mag}
\label{fig:fig3}
\end{figure}    
Finally, we  compare the  kinematical disk masses 
$M_D$    with   those 
derived  from the galaxy specto-photometry:     
$M_{phot}={\mathcal Y}_I \ (B-I) \ L_I$
(let us remind that  ${\mathcal  {Y}}_I$ is 
the average stellar mass-to-light ratio for
a stellar
population 
of color $B-I$.  As a quite general outcome  of the past history of                            
galaxies,  Vazdekis et
al. 1996 found 
$$
{\mathcal {Y}}_I=b \  (B-I)+ c
\eqno(3b)
$$
where $b\sim \ 0.8$ and the value of the 
constant $c$ is irrelevant.  We find:
$$
M_D=(0.96 \pm 0.1) M_{phot}
\eqno(4)
$$ 
(28 d.o.f., see Fig 4), i.e.  dynamical and 
photometric mass estimates statistically  coincide.
Notice  that the  disk mass and I-band  luminosity are not
directly proportional: 
$$
log \  M_D=(0.79 \pm 0.1)\  log L_I +
const
\eqno(5)
$$ 
with the log slope being significantly different from  unity. 
Then, the 
existence of   a relationship
between the  maximum rotation velocity and the  galaxy
luminosity is far from being a   trivial one,  in that  the $I$-band
luminosity  is  not a straightforward measure of the stellar  mass.
  It is  worth to note that in the 3-D space defined 
by the  ($log L_{I}$, $log (M_D/L_I)$, $  B-I$)
coordinate vector, spirals are not randomly distributed, but  
occupy a very thin plane (see Fig. 2).

%%%%%%%%%%%%%%%%%%%%%%%%%%%%%
%%
%%%%%%%%%%%%%%%%%%%%%%%%%%%%%
%%
%%%%%%%%%%
\begin{figure}
\vspace{9.5cm}
\includegraphics{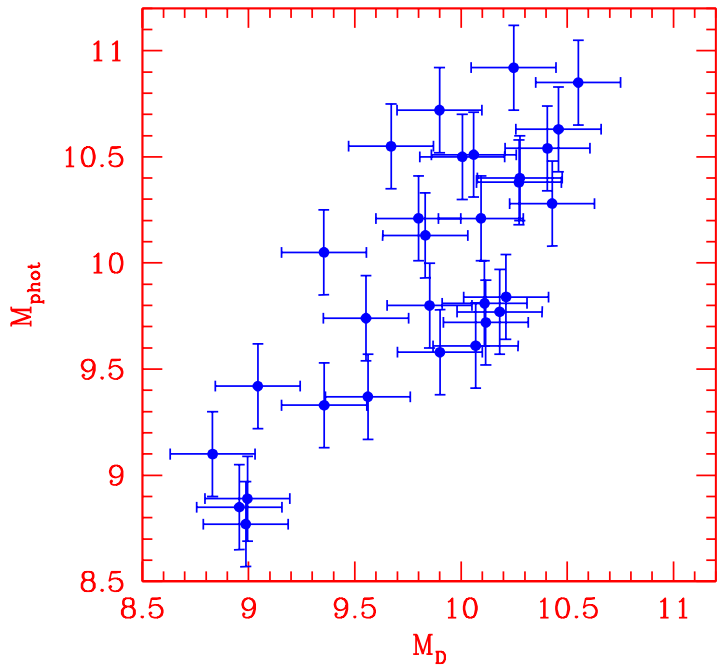}
\caption{ Kinematical mass $M_D$  {\it vs} photometric  disk
masses. Units in $log M_\odot$}
\bigskip
\label{fig:fig4}
\end{figure}

%%%%%%%%%%%%%%%%%%%%%%%%%%%%%
%%
%%%%%%%%%%%%%%%%%%%%%%%%%%%%%
%%
%%%%%%%%%%%
\section {MDO/BH masses. Upper limits}

The very good
spatial resolution   of the RC's of our sample   
plays a crucial role   in the determination of strict and reliable  MDO  
upper 
limits. In  fact, since at small radii
$V\propto R $ and the  mass inside $r_{in}$ scales as    $  r_{in}^3$,  a   spatial 
resolution of   $r_{in} \sim 100pc$  yields to  a mass resolution
   of $\sim 10^6M_\odot$,
given that $V(r_{in}) \sim 10 km/s$.
In detail, the
ratio of the rotational velocities $(V_{MDO}/V_d)^2
 \propto (M_{MDO}/r_{in})^3$,
 implies that  a 
higher spatial resolutions lead  to  stronger limits on 
 $M_{MDO}$ (or easier detection).

We now force the  presence of a central MDO/BH by adopting 
the maximum possible MDO mass compatible with the rotation curve.
We determine the upper mass limits by assuming $f=1$ or $f=f_{max}$ with 
 $f=f_{max}$ the value that brings the model RC  
$1\sigma$ higher than the innermost data. Note that for $f=1$
all the mass inside $r_{in}$ is in the black hole. We typically find:
 $f_{max}=0.7-0.8$ and that, in spite of having an additional parameter,
  the disk+black hole   model performs  
significantly worse than the OD model.

Notice that only in a few cases  $f_{max}\sim 0.2$, furthermore, 
 no result changes if we assume$ f_{max}=1$ for all the objects.
$M_{BH}$ is then
obtained by substituting in equation (2b)  the 
corresponding values of $f$.
These are  shown in Table 1: it is evident that typically 
$M_{MDO}<<10^8 M_{\odot}$ especially at low luminosity.
It is clear that these 
limits are  short of the   mass residing in a accreting BH 
which is  required  to  
power
 high redshifts quasars, thus, 
 spirals do
not host QSO remnants (see also Salucci etal 2000). 
More specifically,  if  central MDO's of
masses 
$\sim 2\ M_{MDO}$ (and 
still $<10^8 M_\odot$)  were actually
present at the center of the late type spirals 
 of our sample,  they would have 
 affected the  inner kpc of the available RC's  
more strongly  than MDO's 
of $\sim 10^9 M_\odot$
affect  the inner  kinematics of ellipticals 
(Magorrian et al., 1998).  
 
%%%%%%%%%%
\begin{figure}
\vspace{8.5cm}
\includegraphics{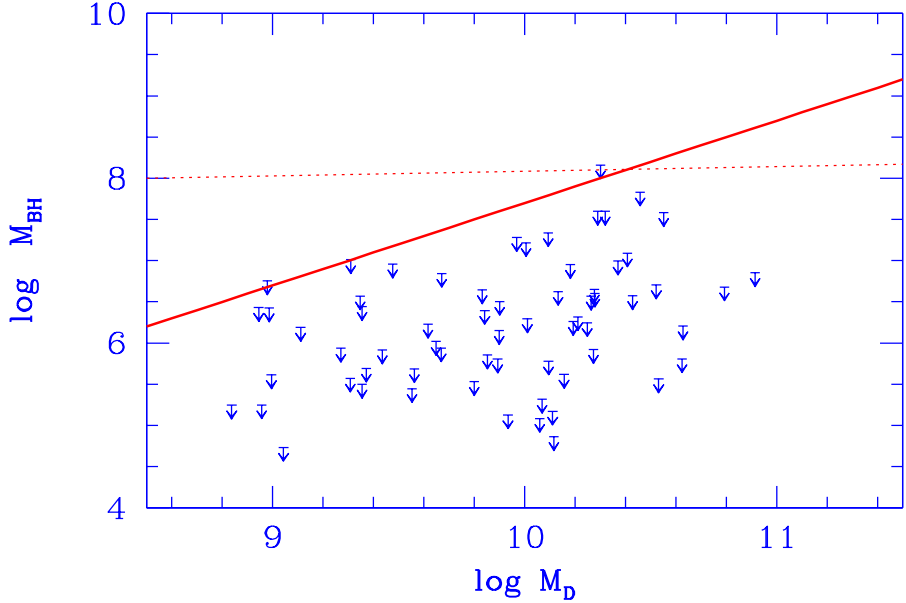}
\caption{ Upper limits BH/MDO masses {\it  vs.} disk
masses. The solid line is the BH/MDO mass {\it  vs.} stellar 
mass of  ellipticals.}
\label{fig:fig5}
\end{figure}
%%%%%%%%%%%%%%%%%%%%%%%%%%%%%

In Fig. 5 we show the relationship  $M_{BH}$ {\it vs } the  stellar
disk mass $M_D$, and, as a reference, we  plot  the
extrapolation to spirals of the elliptical's 
stellar mass-MDO/BH mass relationship  
$M_{MDO}=0.005 \times M_{stars}$ (Magorrian et
al.,1998, Kormendy \& Richstone 1995 ).
We immediately realize, that  in
spirals, the MDO masses, if different from zero, are 
however much smaller than those
detected at
the centers of ellipticals of same stellar luminosity; 
whether this is related  to their less prominent  bulge content or does  
reflect also  a morphological mass  segregation, 
will be  considered elsewhere (Salucci et al 2000). Notice that in the same paper
we perform a  detailed investigation on possible biases occurring
in the present  estimate of the   MDO upper limits.

\section {DISCUSSION} 

Accurate mass models
of the innermost kpc of 83 spirals derived   from    
high  spatial resolution   RC's,  reveal that,
inside this region, the luminous  matter  fully  accounts for
the observed kinematics: a stellar disk of constant mass-to-light ratio (in
some cases in conjunction  
with a spheroid) is virtually the only mass component. 
In this region, with the exception of
very high luminosity galaxy with complex dynamics, 
we find 
$$
V^2(R) \propto   {\sl light(<R)} /R
\eqno(6) 
$$ 
which is indicative of the absence of a sizable  dark component.
Therefore, the only possibility for the presence of a dark halo in the
central regions of spirals 
is that it  conceals itself  below the detection  threshold.

More in detail, inside the innermost
kpc or so, the contribution to the gravitational potential from 
a dark component cannot exceed 10\%; this is  
in disagreement with claims and scenarios   in which a dark 
halo has a major  role at 
any radii. Moreover, in  the  same objects studied here,
the DM, undetected at  $R< 1/3 \  R_D$, begins to  
dominate the mass distribution at  $R\sim 3R_D$ (PSS). Then, the dark
mass inside a giv en  radius $R$ must  increase with $R$
very steeply,  e.g. as in the case of a constant 
density distribution.

In  a DM-free environment,  we have been
able to estimate the disk masses also for objects in which the
global influence of the dark matter  is relevant and to
relate disk colors and  masses across a large range of galaxy 
luminosities. The next step will be to pursue a
coordinated  kinematical and photometric study  to tackle open   
cosmological issues.  
\section{Appendix}
In the appendix we produce the two Tables  of galaxy and
MDO/BH properties.
\nopagebreak
\begin{table*}[hb]
\centering
\begin{tabular}{||lcccc||}
\hline 
${\it name}$&$M_{D}$&B-I&${\mathcal{Y}}$&$L_{I}$ \\
 116-G12       &9.83       &.675              &.51&  9.59\cr
  121-G6       &10.27       &.700            &.72     
&9.84\cr
  143-G10       &8.83       &.225          &-.08     
&8.92\cr
  346-G26      &10.11     &-.125          &.247      &9.82\cr
  347-G28       &9.56     &-.150            &.01     
&9.49\cr
  347-G33      &10.40       &.300          &.23    
&10.30\cr
  347-G34      &10.27       &.025            &-.08    
&10.36\cr
  357-G16       &8.99      &-.350           &-.32     
&9.17\cr
  359-G6        &9.04      &-.100           &-.5     
&9.50\cr
  362-G11       &9.89       &.775             &.11    
&10.10\cr
  374-G8        &8.95       &.000             &.11     
&8.85\cr
  380-G23       &9.67       &.875              &.17     
&9.85\cr
  406-G26      &10.11       &.000            &.30     
&9.81\cr
  418-G1        &9.85       &.050             &.11     
&9.76\cr
  441-G2       &10.18      &-.050             &.35     
&9.81\cr
  468-G23       &9.55       &.000            &-.19     
&9.74\cr
  487-G2       &10.00       &.525             &.14    
&10.08\cr
  488-G54      &10.21      &-.250            &.07    
&10.04\cr
  51-G18       &10.06      &-.075             &.37     
&9.67\cr
  533-G53      &10.06       &.125            &-.30    
&10.41\cr
  54-G21        &9.35       &.375            &-.25     
&9.75\cr
  554-G28       &9.35       &.075             &.12     
&9.27\cr
  556-G23      &10.09       &.200             &.12    
&10.05\cr
  566-G22       &9.90      &-.150             &.14     
&9.70\cr
  60-G25        &8.98      &-.250            &-.08     
&8.97\cr
  84-G10        &9.80       &.050            &-.35    
&10.17\cr
  305-G6       &10.45       &.250             &.13    
&10.43\cr
  476-G15      &10.42      &-.025             &.12    
&10.30\cr
  79-G14       &10.55       &.525              &.33    
&10.43\cr
  79-G3        &10.24       &.800             &.29    
&10.28\cr
\hline  

\end{tabular}

\smallskip

\caption{ (1) Galaxy name (2) Logarithm of the disk
mass in solar units
(3) B-I color (4) Logarithm of the disk mass-to-light ratio (I-band).
(5) I-Band luminosity.}

\end{table*} 

\begin{table*}
\centering
\begin{tabular}{||lccccc||}
\hline
{\it name}&$log M_{BH}$&{\it name}&$log M_{BH}$&{\it
name}&$log \
M_{BH}$\\

     116-g12     &6.6&      418-g8     &6.4&    m-1-2524    
&7.0\cr
      121-g6     &6.7&      436-g3     &6.7&    m-214003    
&6.3\cr
      13-g16     &5.9&      441-g2     &7.0&    m-3-1042    
&5.8\cr
     143-g10     &5.2&     446-g53     &6.8&       n1090    
&5.6\cr
     157-g20     &7.2&     468-g23     &5.4&       n1832    
&8.3\cr
     157-g38     &5.9&     481-g13     &5.6&       n2763    
&7.6\cr
     162-g17     &6.2&      487-g2     &7.2&       n3715    
&8.2\cr
     233-g42     &6.3&     488-g54     &6.3&       n4348    
&6.2\cr 

     249-g16     &5.8&     490-g28     &5.9&        n699    
&6.3\cr
     249-g35     &5.5&     490-g45     &7.3&       n7218    
&7.6\cr 
      265-g2     &6.7&      498-g3     &6.6&       n7339    
&6.6\cr
      302-g9     &6.6&      51-g18     &5.3&        n755    
&5.7\cr
     328-g43     &6.2&     533-g53     &5.1&        ua17    
&5.8\cr
      329-g7     &7.2&      54-g21     &5.5&     269-g19    
&8.3\cr
     346-g26     &4.9&      545-g3     &5.1&      305-g6    
&7.8\cr
     347-g28     &5.7&      545-g5     &6.4&     358-g63    
&6.7\cr
     347-g33     &7.1&     547-g24     &7.0&      36-g19    
&7.7\cr
     347-g34     &5.9&       55-g4     &6.3&     476-g15    
&6.6\cr
      35-g18     &7.7&      550-g7     &5.6&     545-g11    
&8.3\cr
     357-g16     &5.6&     554-g28     &6.4&     547-g31    
&7.4\cr
      359-g6     &4.7&     556-g23     &7.3&     582-g12    
&7.7\cr
     362-g11     &6.2&     563-g14     &7.3&      79-g14    
&7.6\cr
      374-g8     &5.2&     566-g22     &6.5&       79-g3    
&6.2\cr
     380-g23     &6.8&     575-g53     &6.0&      i96099    
&6.9\cr
     406-g26     &5.2&      58-g30     &6.6&    m-222025    
&8.6\cr
     410-g19     &7.0&      60-g25     &6.4&       n4705    
&7.2\cr
      418-g1     &5.9&      84-g10     &5.5&        n697    
&8.4\cr
       I407     &7.&        I96099     &6.9&&   \cr

\hline
\bigskip
\end{tabular}
\caption{ BH/MDO upper limits for the galaxies of the sample, in units of $log M_\odot$ }
\end{table*}

\end{document}